\documentclass[aps,prd,10pt,twocolumn,nofootinbib,superscriptaddress]{revtex4-2}
\usepackage{epsfig}
\usepackage{bm}
\usepackage{latexsym}
\usepackage{natbib}
\usepackage{url}
\usepackage[T1]{fontenc}
\usepackage{lmodern}
\usepackage{color}
\usepackage{amsfonts,amssymb,amsmath}
\usepackage{graphicx,epsfig} 
\usepackage{psfrag}
\usepackage{subfigure}
\usepackage[citecolor = ProcessBlue]{hyperref}
\hypersetup{colorlinks=true}
\usepackage{mathtools}
\usepackage{enumitem}
\usepackage{float}
\usepackage{mathrsfs}
\usepackage[dvipsnames]{xcolor}
\usepackage[utf8]{inputenc}
\usepackage{comment}
\allowdisplaybreaks
\labelformat{section}{Section~#1} 
\labelformat{equation}{Eq.~(#1)} 
\labelformat{figure}{Fig.~#1} 
\labelformat{subfigure}{Fig.~\thefigure#1} 
\labelformat{table}{Table~#1} 
\begin{document}

\title{Manifestation of  quantum entanglement between harmonic oscillators in de Sitter}

\author{Sumanta Chakraborty}
\email{tpsc@iacs.res.in}
\affiliation{School of Physical Sciences, Indian Association for the Cultivation of Science, Kolkata 700032, India}

\author{Anupam Mazumdar}
\email{anupam.mazumdar@rug.nl}
\affiliation{Van Swinderen Institute for Particle Physics and Gravity, University of Groningen, 9747AG Groningen, the Netherlands}

\author{Ritapriya Pradhan}
\email{rxu9em@virginia.edu}
\affiliation{Department of Physics, University of Virginia, P.O. Box 400714, Charlottesville, VA 22904-4714, USA}
\affiliation{School of Physical Sciences, Indian Association for the Cultivation of Science, Kolkata 700032, India}

\begin{abstract}
Two harmonic oscillators interacting through the exchange of a quantum field leads to non-zero entanglement between the two, which is absent for classical interaction. In this work, we determine the entanglement between two such harmonic oscillators living in an expanding universe. It turns out that, if the oscillators are within the Hubble horizon, with their frequencies are comparable to the rate of expansion of the universe, the entanglement is non-zero and significant. While, for oscillators outside the Hubble horizon, with oscillation frequencies much higher than the expansion rate, the entanglement is negligibly small. 
\end{abstract}

\maketitle

\section{Introduction}

Entanglement in quantum information theory has brought many insights into numerous seemingly unrelated areas of physics, in particular with a growing interest in theory and experiment in the context of entanglement entropy~\cite{Horodecki1996}. Quantum interaction inevitably yields quantum entanglement, and it is now evident in  ordinary local quantum field theories (QFTs) \cite{Calabrese2004, Casini:2009sr, Metlitski:2009iyg, Hertzberg2013, Seki2015, Witten:2018zxz, Casini:2022rlv, Headrick2018TASILO, Hollands:2017dov, Fedida2023,Hertzberg2013,
Fedida:2024dwc}, holographic quantum field theories, see~\cite{Casini:2011kv,Rangamani:2016dms}; in the context of gravity ~\cite{Bekenstein:1981,Sorkin:1986,Srednicki:1993im,tHooft:1993dmi}; in scattering processes of particles~\cite{Balasubramanian:2011wt,Peschanski2016,Peschanski2019,Bose2020}, non-local quantum field theory (NLQFT) \cite{Landry2024,Vinckers:2023grv}.

Entanglement has also featured in devising an experimental test to witness the quantum nature of gravitational interaction with matter in a lab
\cite{Bose:2017nin,ICTS,PhysRevLett.119.240402,
Marshman:2019sne, Carney_2019,Bose:2022uxe,Biswas:2022qto,Carney:2021vvt,Elahi:2023ozf,Hanif:2023fto,Chakraborty:2023kel}; hence put forward the experimental test to witness the hypothetical massless spin-2 quanta known as graviton~\cite{Bose:2017nin,Marshman:2019sne,Danielson2022}.

A natural question arises: how would entanglement between two quantum systems evolve in an expanding Universe? In the cosmological context, this question has been studied vigorously, albeit at the level of understanding the decoherence (the quantum to classical transition) of the cosmic microwave background radiation~\cite{Halliwell:1987es,Brandenberger:1990bx, Albrecht:1992kf, Polarski:1995jg,Calzetta:1995ys,
Lesgourgues_1997,Kiefer:1998qe,NAMBU199211,BARVINSKY1999374,Kiefer:2006je,Martineau:2006ki,Kiefer:2006je,
Martin:2012pea,Martin:2018zbe,Martin:2021znx,Martineau:2006ki,Brandenberger:1990bx,Burgess_2024,Burgess_2025,Hoover:2008,Weenink:2011dd,Colas_2023,Brahma:2020zpk,Lopez:2025,sano2025, Raveendran:2023dst, Rajeev:2019okd, Rajeev:2017uwk, Rajeev:2017smn}. 

The aim of the current paper is different; here, we wish to set up two quantum systems, to be precise, harmonic oscillators in a de Sitter background and study how entanglement unfolds between the two oscillators interacting through a scalar field. This scalar field is assumed to be massless, while we assume that the two harmonic oscillators possess similar mass $m$ and frequency $\Omega$. Given the above setup, we will show that if the oscillators are separated by a distance, $d(\ll H^{-1})$, i.e., the oscillators are within the Hubble horizon of the de Sitter universe, located at $H^{-1}$, and oscillating at a rate comparable to the expansion rate of the universe, e.g., $\Omega \sim H$, then the entanglement is large and non-zero at late times. While for oscillators outside the Hubble horizon, with $d\gg H^{-1}$ and oscillating at a rate much faster than the expansion rate of the universe, $\Omega \gg H$, the entanglement is still non-vanishing, but negligible. 

We will begin with a discussion of our setup in \ref{timedependent_entagle}. Then, in \ref{hamiltonian}, we will study the interaction between the two harmonic oscillators via a scalar field in the de Sitter background and determine the interaction Hamiltonian. Finally, we will examine the evolution of the entanglement between these oscillators, arising out of the above interaction Hamiltonian, for the expanding de Sitter background, in \ref{entanglement_dS} and then we conclude.

\emph{Notations and Conventions:} We use mostly positive signature convention, such that the flat spacetime metric in the Cartesian coordinates become, $\eta_{\mu \nu}=\textrm{diag.}(-1,+1,+1,+1)$. The Greek indices $\alpha,\beta,\mu,\cdots$ denotes spacetime indices and the Roman indices $i,j,k,\cdots$ describe spatial coordinates.

\section{Entanglement in a time-dependent scenario: The basic setup}\label{timedependent_entagle}

We consider two identical harmonic oscillators, denoted as `A' and `B', having mass $m$ and frequency $\omega$, located at $x=\pm d/2$, respectively. Incorporation of quantum effects lead to fluctuations of these harmonic oscillators around their respective classical positions. Therefore, one may express the position operator associated with these two harmonic oscillators as, 
\begin{equation}\label{setup1}
\hat{x}_A=-\frac{d}{2}+\delta \hat{x}_A\;;\;\hat{x}_B=-\frac{d}{2}+\delta \hat{x}_B~.
\end{equation}
Note that $\delta \hat{x}_A$ and $\delta \hat{x}_B$ are the quantum fluctuations associated with this system. In the absence of any interaction, the Hamiltonian associated with this system is given by, 
\begin{equation}
\hat{H}=\hat{H}_A+\hat{H}_B~;
\quad
\hat{H}_{A,B}=\frac{1}{2m}\hat{p}_{A,B}^2+\frac{m\Omega^2}{2}\delta \hat{x}_{A,B}^2~.
\end{equation}
The gravitational interaction between these two masses, which have a quantum origin, will perturb the above Hamiltonian. The perturbation is going to manifest as an entanglement between these two harmonic oscillators, leading to the macroscopic superposition of masses. Therefore, the existence of such an entanglement in this two-harmonic-oscillator system is a tell-tale sign of the quantum nature of gravitational interaction. The entanglement between these harmonic oscillators due to quantum gravitational interaction has already been studied in flat spacetime \cite{Bose:2022uxe}. Subsequently, this has been generalised in \cite{Chakraborty:2023kel} to include additional scalar interactions due to frame transformations, and it was demonstrated that entanglement can also serve as a probe to distinguish between Einstein and Jordan frames. In this work, we will explore the fate of entanglement in a time-dependent situation, both for gravitational and scalar interactions between harmonic oscillators, in a time-dependent background spacetime, specifically de Sitter.  

Unlike flat spacetime, in a time-dependent background, the perturbations are also time-dependent. Hence, the time-independent perturbation theory no longer applies, and we should use the time-dependent perturbation theory to determine the entanglement. It is expected that the entanglement is also time-dependent. Adding a time-dependent perturbation to the original unperturbed Hamiltonian of this two-harmonic-oscillator system, we express the Hamiltonian as $\hat{H}=\hat{H}_A+\hat{H}_B+\hat{H}_{AB}$. Due to the perturbation, the initial state of the two-harmonic oscillator system will change. We assume that initially, both the harmonic oscillators are in the ground state, such that, 
\begin{equation}
\left|\psi_{\mathrm{i}}\right\rangle=|0\rangle_{A}|0\rangle_{B}~.
\end{equation}
In the presence of the time-dependent interaction, the final state will be time-dependent, which can be expanded in terms of the basis states as 
\begin{equation}
\left|\psi_{\mathrm{f}}\right\rangle \equiv \frac{1}{\sqrt{\mathcal{N}}} \sum_{n, N} C_{n N}(t)|n\rangle_{\rm A}|N\rangle_{\rm B}~,
\end{equation}
where $\mathcal{N}$ is the normalisation factor. The coefficients $C_{nN}$ are chosen in such a manner that $C_{00}=1$, while the other coefficients are time-dependent and are obtained by employing the time-dependent perturbation theory. For our purpose, linear order perturbation theory suffices, since the strength of the perturbation is small due to the smallness of the coupling constants, be it gravity or scalar. For that we have the following expression for these coefficients, see e.g. \cite{Bose:2022uxe},
\begin{equation}\label{gen_excite}
C_{n N}(t)= \frac{-i}{\hbar}\int_{t_0}^t dt' e^{i\omega_{nN,00}t'} \,_{A}\langle n|_{\rm B}\langle N| \hat{H}_{A B}(t')|0\rangle_{\rm A}|0\rangle_{\rm B}~,
\end{equation}
where $\hbar\omega_{nN,00}$ is the energy difference between the states $|0\rangle_{\rm A}|0\rangle_{\rm B}$ and $|n\rangle_{\rm A}| N\rangle_{\rm B}$ under the unperturbed Hamiltonian, and $t_0$ is the time from when the perturbation is switched on. 

In general, the state $\left|\psi_{\mathrm{f}}\right\rangle$ cannot be written as a product of states associated with the `A' and `B' harmonic oscillators, and hence is not a pure state. To quantify the entanglement, we define a measure called concurrence, which is given by
\cite{Wootters1998,Hill1997,Rungta_2001}
\begin{equation}\label{concurrence}
C \equiv \sqrt{2\left(1-\mathrm{tr}\left[\hat{\rho}_{\rm A}^{2}\right]\right)}~,
\end{equation}
where the density matrix, $\hat{\rho}_{\rm A}$ can be computed by tracing away the states associated with the `B' harmonic oscillator, yielding
\begin{equation}\label{densityA}
\hat{\rho}_{A}=\sum_{N}\,_{B}\langle N| \psi_{\mathrm{f}}\rangle\langle\psi_{\mathrm{f}}|N\rangle_{\rm B}~.
\end{equation}
If the state is pure, then $\hat{\rho}_{A}^2=\hat{\rho}_{A}$ and hence $\mathrm{tr}\left[\hat{\rho}_{A}^2\right]=1$. Therefore, it follows that the concurrence, as defined in \ref{concurrence}, identically vanishes. The greater the value of the concurrence, the more entangled the two harmonic oscillators are. In the subsequent sections, we will first determine the interaction Hamiltonian for gravity and scalar in the dS background, and then find out the concurrence to assess whether the quantum nature of gravity can be probed in a cosmological setting.  

\section{Real Scalar field on de Sitter background: Interaction hamiltonian}\label{hamiltonian}

In the flat spacetime, the interaction between two massive objects is mediated by gravitons in the Einstein frame. However, in the Jordan frame, which involves non-minimal coupling between gravity and a scalar degree of freedom, a part of the interaction between two massive bodies is also mediated by scalars. Thus, for generality, we also consider both gravity and scalar mediation between these harmonic oscillators on the dS background.  

In this section, we will discuss the dynamics of a scalar field on the dS background and hence determine its contribution to the interaction Hamiltonian. The starting point is the free field action for the scalar field
\begin{equation}
\label{4.1}
\mathcal{A}=-\frac{1}{2} \int d^4 x \sqrt{-g} \left[ g^{\alpha \beta} \nabla_\alpha \phi \nabla_\beta \phi + m_{\phi}^2 \phi^2 \right]~,
\end{equation}
where $m_{\phi}$ is the mass of the scalar field and $g_{\alpha \beta}$ is the spacetime metric, which for dS spacetime in the cosmological coordinates is given by, $g_{\alpha \beta}=\textrm{diag.}(-1,a^{2}(t)\delta^{i}_{j})$, with $a(t)=e^{Ht}$. Here, $H^{2}=\textrm{constant}=(\Lambda/3)$. For our purpose, it will be convenient to transform to the conformal coordinates, with the conformal time $\eta$ being defined as,
\begin{equation}
\label{4.3}
\eta\equiv\int^{t}\frac{dt'}{a(t')}~;
\qquad
a(\eta)=-\frac{1}{H\eta +1}~,
\end{equation}
where, $\eta \in (-\infty,-H^{-1})$. In this coordinate system involving conformal time, the metric of the dS spacetime can be expressed as, $g_{\alpha\beta}= a^2(\eta)\eta_{\alpha \beta}$. Besides, it is instructive to rescale the scalar field as $\chi=a(\eta)\phi$, such that \ref{4.1} reduces to the following form, 
\begin{equation}
\label{4.5}
\mathcal{A}=\frac{1}{2} \int d^3\mathbf{x} d\eta\; \left[\chi'^2 - (\nabla \chi )^2 - \left( m_{\phi}^{2}a^2-\frac{a''}{a} \right) \chi^2 \right]~,
\end{equation} 
where `prime' denotes derivative with respect to the conformal time $\eta$. This action resembles that of a scalar field moving on a flat background, however with a time-dependent mass. Variation of the above action with respect to $\chi$ yields the following equation of motion,
\begin{equation}
\label{4.6}
\chi ''-\nabla^{2}\chi+\left(m_{\phi}^{2}a^{2}-\frac{a''}{a}\right)\chi=0~.
\end{equation}
Now we shall quantize this scalar field. Following the canonical prescription, given the action in \ref{4.5}, we first determine the conjugate momenta to the scalar field, which turns out to be $\Pi = \chi'$, and then impose the canonical quantization rules on the field and its conjugate momenta. For this purpose, and following the spatial isotropy of the dS spacetime we express the scalar field in terms of the Fourier modes as, 
\begin{equation}
\label{4.11}
\hat{\chi}(\mathbf{x},\eta)=\frac{\mathcal{G}}{\sqrt{2}}\int \frac{d^3 \mathbf{k}}{(2 \pi)^{\frac{3}{2}}} \left(\hat{a}_{\mathbf{k}}v^*_{\mathbf{k}}(\eta)e^{i\mathbf{k}\cdot \mathbf{x}}+\hat{a}^{\dagger}_{\mathbf{k}}v_{\mathbf{k}}(\eta)e^{-i\mathbf{k}\cdot \mathbf{x}}\right)~.
\end{equation}
Here, we have used the fact that, $\chi(\textbf{x}, \eta)$ is a real field, and hence we must have 
$\chi_\textbf{k}^{*}=\chi_{-\textbf{k}}$. The mode functions $v_{\mathbf{k}}(\eta)$ satisfies the following equation,
\begin{equation}
\label{4.8}
v_{\textbf{k}}'' + \omega_{\mathbf{k}}^{2}(\eta) v_{\textbf{k}} = 0~, 
\quad 
\omega_{\mathbf{k}}^2(\eta)=\mathbf{k}^2+m_{\phi}^2a^2(\eta)-\frac{a''}{a}~,
\end{equation}
along with the normalization condition for the mode functions: $\mathrm{Im}(v'_{\mathbf{k}} v^{*}_{\mathbf{k}})=1$. Besides, the creation and the annihilation operators satisfy the following commutation relations: $[ \hat{a}_\mathbf{k}, \hat{a}^\dagger_\mathbf{k'} ] = \delta (\mathbf{k}-\mathbf{k'})$, while the other commutation relations between the creation and the annihilation operators vanish.

For the dS background, the differential equation satisfied by the mode functions $v_{\mathbf{k}}(\eta)$ can be solved, with the initial condition coming from choosing our vacuum to be the Bunch-Davies vacuum, which yield,
\begin{align}
\label{4.15}
v_{\mathbf{k}}(\eta)&=\sqrt{\frac{\pi|H\eta + 1|}{2H}}\Big[J_{n}\left(\frac{k}{H}|H\eta+1|\right)
\nonumber
\\
&\qquad \qquad \qquad -iY_{n}\left(\frac{k}{H}|H\eta+1|\right)\Big]~,
\end{align}
where, $k\equiv |\mathbf{k}|$, and $n= \sqrt{(9/4)-(m_{\phi}^2/H^2)}$, such that for massless field $n=(3/2)$. In what follows we will work with a massless scalar field, for which, using the decomposition of the respective Besel functions in terms of elementary functions, we obtain,
\begin{align}
\label{4.16}
v_{\mathbf{k}}(\eta)=\frac{1}{\sqrt{k}}\left(\frac{ie^{-ix}}{x}-e^{-ix}\right)~;
\quad 
x\equiv \frac{k}{H}|H\eta+1|~.
\end{align}
Having obtained the mode functions, i.e., knowing the quantity $v_{\mathbf{k}}(\eta)$ explicitly, the free field Hamiltonian can be expressed in terms of the creation and annihilation operators, along with the mode function as, 
\begin{equation}
\label{5.7}
\hat{H}=\int \frac{d^3\mathbf{k}}{4}\left[\hat{a}_{\mathbf{k}}\hat{a}_{-\mathbf{k}} F^*_{\mathbf{k}}+\textrm{h.c.}
+\left(2\hat{a}^{\dagger}_{\mathbf{k}}\hat{a}_{\mathbf{k}}+\delta^{(3)}(0)\right)E_{\mathbf{k}}\right]~,
\end{equation}
where, $\textrm{h.c.}$ denotes hermitian conjugate of the first term involving $F_{\mathbf{k}}$, as the Hamiltonian must be hermitian, and the quantities $E_{\mathbf{k}}$ and $F_{\mathbf{k}}$ are defined as,
\begin{eqnarray}
E_{\mathbf{k}}&=&|v'_{\mathbf{k}}(\eta)|^2+ \omega^2_{\mathbf{k}}(\eta)|v_{\mathbf{k}}|^2~,
\\ 
F_{\mathbf{k}}(\eta)&=& v'^2_{\mathbf{k}} + \omega^2_{\mathbf{k}}(\eta)v_{\mathbf{k}}^2~.
\end{eqnarray}
The next step is to use the expressions for the mode functions, as well as those of the free field Hamiltonian, to determine the interaction Hamiltonian between the two harmonic oscillators.  

For this purpose, we start by writing down the interaction between the matter (consisting of two harmonic oscillators) and the scalar field. The matter sector is described by the energy momentum tensor $T_{\mu \nu}$ and the scalar field is by $\phi$, such that the interaction between them is described by \cite{Chakraborty:2023kel}
\begin{equation}
\label{5.1}
\mathcal{A}_{\rm int}=\int d^4 x \sqrt{-g} \; \phi\; (T_{\mu \nu}g^{\mu \nu})~.
\end{equation}
In the conformal coordinates, the above interaction in terms of the rescaled scalar field $\chi$ becomes,
\begin{equation}
\label{5.2}
\mathcal{A}_{\rm int}=\int d^3\mathbf{x}d\eta\; a(\eta)\;\chi\left(T_{\mu \nu}\eta ^{\mu \nu}\right)~,
\end{equation}
where, $\phi=a(\eta)\chi$, with $a(\eta)$ being the scale factor and $\eta$ being the conformal time, defined in \ref{4.3}. Note that, we have kept $T_{\mu \nu}$ to be the same in both of the above expressions, but it is understood that the expression of $T_{\mu \nu}$ as a function of $\eta$ will be different from the corresponding expression in terms of the cosmological time.

So far, our discussion regarding the interaction between these harmonic oscillators through the exchange of a scalar field was classical. Moving onto the quantum domain, we note that since the matter consists of two harmonic oscillators located at $\mathbf{x}_{\rm A}$ and $\mathbf{x}_{\rm B}$, respectively, the operator form of the energy-momentum tensor of this system, is given by 
\begin{equation}
\label{5.3}
\hat{T}_{\mu\nu}(\mathbf{x})=\frac{\hat{p}_{\mu}\hat{p}_{\nu}}{\sqrt{-g}p^0}\left[\delta^{3}(\hat{\mathbf{x}}-\hat{\mathbf{x}}_A)+\delta^{3}(\hat{\mathbf{x}}-\hat{\mathbf{x}}_B)\right]~.
\end{equation}
Here, $\hat{p}_{\mu}$ is the four-momentum operator, whose classical expression is given by $p_{\mu}=mu_{\mu}$, for both the identical harmonic oscillators, and $\hat{x}_{A,B}$ denote position operators, whose classical realization denotes the positions of the oscillators. Assuming that the oscillations are happening in the $x$ direction alone, it follows that only the $\hat{T}_{00}$, $\hat{T}_{01}$ and $\hat{T}_{11}$ components of the energy-momentum tensor contributes. Hence the interaction hamiltonian in the conformal coordinate system reads,
\begin{widetext}
\begin{eqnarray}
H^{(\eta)}_{\mathrm{int}}
=-\frac{\mathcal{G}a(\eta)}{\sqrt{2}}\int d^{3}\mathbf{k}\left[\hat{a}^{\dagger}_{\mathbf{k}}v_{\mathbf{k}}(\eta)\left\{-\hat{T}_{00}(\mathbf{k})+\hat{T}_{11}(\mathbf{k})\right\}+\hat{a}_{\mathbf{k}}v^*_{\mathbf{k}}(\eta)\left\{-\hat{T}^{\dagger}_{00}(\mathbf{k})+\hat{T}^{\dagger}_{11}(\mathbf{k})\right\}\right]~,
\label{inthamiltonnew}
\end{eqnarray}
\end{widetext}
where, the energy-momentum tensor components in the Fourier space are obtained by multiplying $T_{\alpha \beta}(\mathbf{x})$ by $e^{-i\mathbf{k}\cdot\mathbf{x}}$ and then integrating over the spatial volume with $(2\pi)^{-3/2}$ factor, which yields.
\begin{align}
\hat{T}_{\alpha \beta}(\mathbf{k})&=\int \frac{d^3\mathbf{x}}{(2\pi)^{3/2}}\hat{T}_{\alpha \beta}(\mathbf{x})e^{-i\mathbf{k}\cdot\mathbf{x}}
\nonumber
\\
&=\frac{\hat{p}_{\alpha}\hat{p}_{\beta}}{\sqrt{-g}p^{0}}\left(\frac{e^{-i\mathbf{k}\cdot\mathbf{x}_{A}}+e^{-i\mathbf{k}\cdot\mathbf{x}_{B}}}{(2\pi)^{3/2}}\right)~.
\end{align}
The above provides the interaction term between the harmonic oscillators, whose information are in the components $\hat{T}_{\mu \nu}$ and the mode functions $v_{\mathbf{k}}(\eta)$, which incorporates both the scalar field as well as the background de Sitter spacetime. Since our interest is in the interaction between the two quantum oscillators, it can be obtained by tracing out the scalar degree of freedom which is mediating the interaction. This achieved by the following result from perturbation theory: 
\begin{equation}
\label{5.4}
\hat{H}^{(\eta)}_{\rm AB}=\int d^3\mathbf{k}\frac{\,_{\phi}\langle 0|\hat{H}^{(\eta)}_{\mathrm{int}}|\mathbf{k}\rangle_{\phi}\,_{\phi}\langle \mathbf{k}|\hat{H}^{(\eta)}_{\mathrm{int}}|0\rangle_{\phi}}{\,_{\phi}\langle \mathbf{k}|\mathbf{k} \rangle_{\phi} \left(\mathcal{E}_0-\mathcal{E}_{\mathbf{k}}\right)}~,
\end{equation}
where, $|0\rangle_{\phi}$ denotes the vacuum state associated with the scalar field and $|\mathbf{k}\rangle_{\phi}$ correspond to an excited state with a scalar quantum with momentum $\mathbf{k}$, i.e., $|\mathbf{k}\rangle_{\phi} \propto \hat{a}_{\mathbf{k}}^{\dagger}|0\rangle_{\phi}$. Moreover, the energy terms $\mathcal{E}_{0}$ and $\mathcal{E}_{\mathbf{k}}$ appearing in the denominator of the above expression are defined as, $\mathcal{E}_{0}=\,_{\phi}\langle 0|\hat{H}|0\rangle_{\phi}$, and $\mathcal{E}_{\mathbf{k}}=\,_{\phi}\langle \mathbf{k}|\hat{H}|\mathbf{k}\rangle_{\phi}$, where $\hat{H}$ is the free field Hamiltonian for the scalar field on de Sitter. Given the interaction Hamiltonian in \ref{inthamiltonnew}, we obtain the following result, 
\begin{equation}
\label{5.5}
\,_{\phi}\langle \mathbf{k}|\hat{H}^{(\eta)}_{\mathrm{int}}|0\rangle_{\phi}\propto v_{\mathbf{k}}(\eta)\left[-\hat{T}_{00}(\mathbf{k})+\hat{T}_{11}(\mathbf{k})\right]~,
\end{equation}
whose complex conjugation will yield the other term appearing in the expression for $\hat{H}_{\rm I}$. Note that both of these terms will have an overall normalization factor, which will be cancelled by the factor $\,_{\phi}\langle \mathbf{k}|\mathbf{k} \rangle_{\phi}$ in the denominator.

The energy terms in the denominator of \ref{5.4} can be determined by using the expression of the Hamiltonian for the scalar field in \ref{5.7}. From which, it is clear that $\mathcal{E}_{0}$ is proportional to the integral of $E_{\mathbf{k}}$ over the Fourier momentum along with a factor involving $\delta^{3}(0)$, which will be cancelled by $\mathcal{E}_{\mathbf{k}}$, yielding, the following result for the difference:
\begin{eqnarray}
\label{5.11}
\mathcal{E}_0-\mathcal{E}_{\mathbf{k}}=-\frac{1}{2}E_{\mathbf{k}}=-\frac{1}{2}\Big[|v'_{\mathbf{k}}(\eta)|^2+ \omega^2_{\mathbf{k}}(\eta)|v_{\mathbf{k}}|^2\Big]~. 
\end{eqnarray}
We now have simplified the expressions for all the possible terms in the interaction hamiltonian $\hat{H}^{(\eta)}_{\rm AB}$ between the two harmonic oscillators. Therefore, using \ref{5.5} and \ref{5.11}, in \ref{5.4} we obtain,
\begin{equation}
\label{5.12}
\frac{\hat{H}^{(\eta)}_{\textrm{AB}}}{a^2(\eta)}=\mathcal{G}^{2}\int d^3\mathbf{k}\frac{\left(\hat{T}_{00}^\dagger \hat{T}_{11}-\hat{T}_{00}^\dagger \hat{T}_{00}+\hat{T}_{11}^\dagger \hat{T}_{00}-\hat{T}_{11}^\dagger \hat{T}_{11}\right)}{|v'_{\mathbf{k}}(\eta)|^{2}|v_\mathbf{k}|^{-2}+ \omega^2_{\mathbf{k}}(\eta)} ~.
\end{equation}
To obtain a closed form expression for the interaction Hamiltonian between two harmonic oscillators in dS, we shall first go to non-relativistic limit and will only consider the $\hat{T}^\dagger_{00}\hat{T}_{00}$ term, with $\hat{T}_{00}=\hat{p}^{0}(2\pi)^{-3/2}(e^{-i\mathbf{k}\cdot\mathbf{x}_{A}}+e^{-i\mathbf{k}\cdot\mathbf{x}_{B}})$. Even though it is straightforward to include the other components of the energy-momentum tensor, and keep those terms as a post-Newtonian expansion over and above the Newtonian expansion, since our aim is to capture the essential modification to the interaction Hamiltonian due to the dS background, so it suffices for the moment to take a non-relativistic limit. Further, in this case, the time is no longer an operator valued quantity and hence we can replace $p^{0}$ by its classical value $m(d\eta/dt)=(m/a)$, where $m$ is the mass of the oscillator and $a(\eta)$ is the scale factor. Next we use \ref{4.16} and \ref{5.3}, in \ref{5.12}, and perform the angular integral, after that we do a change of variable from $|\mathbf{k}|\equiv k$ to $x \equiv k|\eta + 1/H|$, and finally arrive at the following expression for the Hamiltonian (ignoring self-energy terms)
\begin{align}
&\frac{\hat{H}^{(\eta)}_{AB}}{a^{2}(\eta)}
=\frac{4\pi i\mathcal{G}^{2}\left(p^{0}\right)^{2}}{(2\pi)^{3}}\frac{1}{|\mathbf{\hat{x}}_{A}-\mathbf{\hat{x}}_{B}|}
\nonumber
\\
&\quad \times\int_{-\infty}^{\infty} dx \frac{x(1+x^{2})}{(2x^{4}-2x^{2}-1)}\exp\left(\frac{iHx|\mathbf{\hat{x}}_{A}-\mathbf{\hat{x}}_{B}|}{|1+H\eta|}\right)\,.
\end{align}
The above integral in the interaction Hamiltonian has four singular points, which are the roots of the equation $2x^{4}-2x^{2}-1=0$, and are located at, $x^{2}=-(\sqrt{3}-1)/2$ and at $x^{2}=(\sqrt{3}-1)/2$. Thus two roots appear on the real axis and two on the imaginary axis. 
At this point we have to choose an appropriate contour to perform the integral, and since all the poles are simple poles, we use the techniques of contour integration. Notice if we do not avoid the poles on the real axis and include them in our contour, that will give rise to an oscillatory part to our interaction Hamiltonian, by residue theorem. But this does not make sense physically, since the proper distance between the two oscillators is increasing with time, we actually expect a decaying Hamiltonian rather than an oscillatory one. So to avoid both the poles on the real axis and close the contour in the upper half plane, we therefore choose the contour used for computation of the retarded Green's function in the problem of radiation from a charge. From that we arrive at the following interaction Hamiltonian in conformal coordinates between the harmonic oscillators:
\begin{equation}
\hat{H}^{(\eta)}_{AB}=\frac{\sqrt{3}m^2\mathcal{G}^{2}}{8\pi|\hat{\bf x}_A-\hat{\bf x}_B|}\exp\left(ia(\eta)x_{0i}\frac{|\hat{\bf x}_{\rm A}-\hat{\bf x}_{\rm B}|}{H^{-1}}\right)~,
\end{equation}
with, $x_{0i}$, being the root of the quadratic equation $x^{2}=-(\sqrt{3}-1)/2$, on the positive imaginary axis. Since in the dS universe, or for that matter, in any expanding universe, the cosmological time is the preferred time coordinate, it is useful to express the interaction Hamiltonian in that coordinate. This yields, 
\begin{equation}
\label{5.14}
\hat{H}^{(t)}_{\rm AB}=\frac{\sqrt{3}m^2}{8\pi a(t)|\hat{\bf x}_A-\hat{\bf x}_B|}\exp\left(-a(t)|x_{0i}|\frac{|\hat{\bf x}_{\rm A}-\hat{\bf x}_{\rm B}|}{H^{-1}}\right)~.
\end{equation}
Let us point out some features of this interaction Hamiltonian --- (a) this Hamiltonian reduces to the Newtonian $(1/r)$ form in the limit $H\rightarrow 0$. Notice, in this limit, $a(t)\rightarrow 1$ and the argument of exponential vanishes. Therefore, the flat background result is obtained, modulo an overall negative sign. (b) At late times the interaction vanishes exponentially, as we expect, since the proper distance grows sufficiently large. Notice the argument of the exponential is the ratio of proper distance between the oscillators to the horizon radius, which is the natural length scale associated with the expansion of the spacetime in dS universe. We will now focus on the evolution of entanglement between the harmonic oscillators through the exchange of a scalar. 

\begin{figure*}
    \centering
    \includegraphics[width = 15cm, height = 9.5cm]{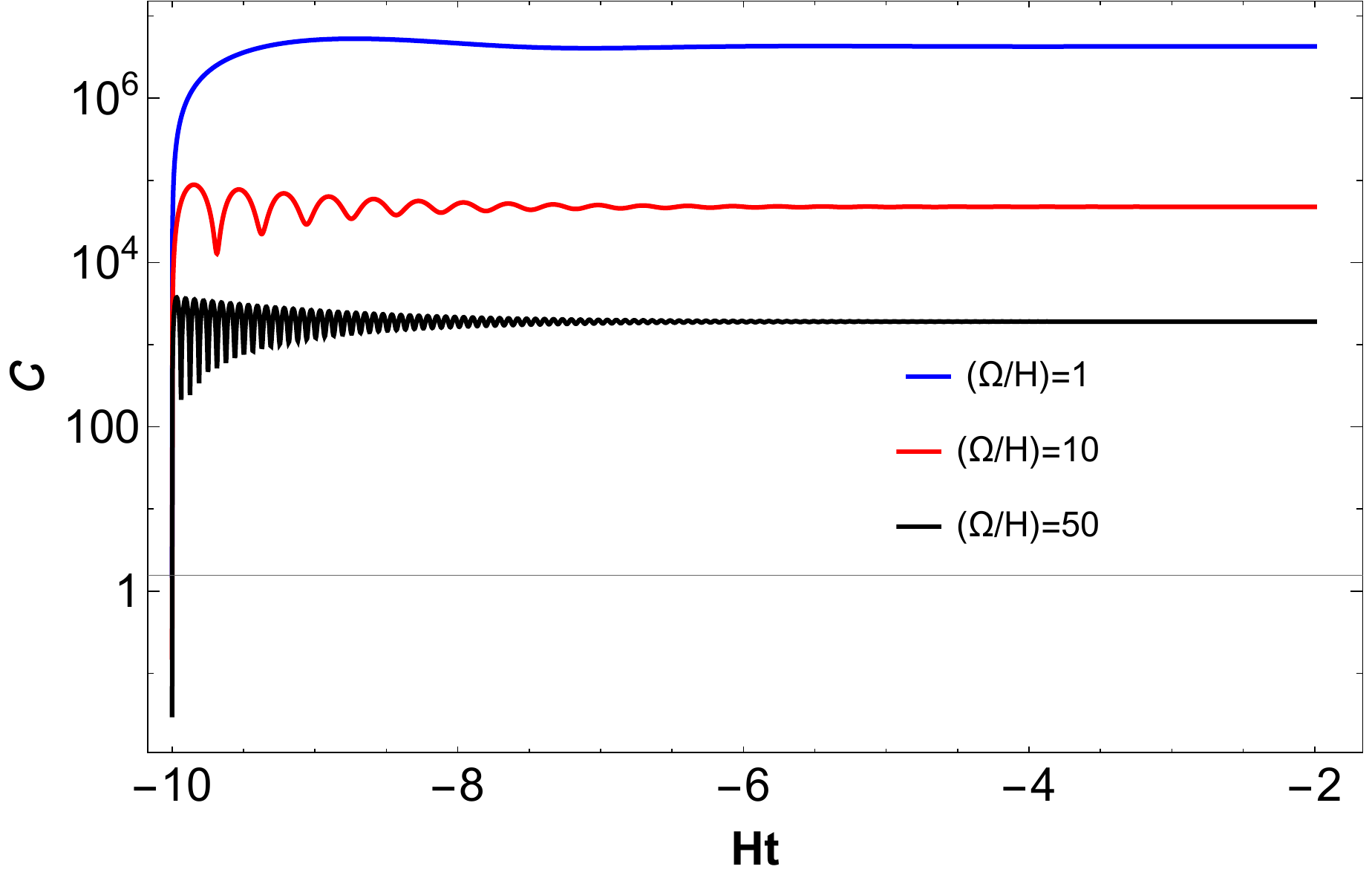}
    \caption{Variation of the rescaled concurrence $\mathcal{C}$ with dimensionless time $Ht$ has been plotted here. Different curves are for three different choices of the dimensionless frequency parameter, $(\Omega/H)=1$, $(\Omega/H)=10$ and $(\Omega/H)=50$, with a fixed dimensionless distance $dH=0.1$. The concurrence saturates to a non-zero and constant value at late times, while experience oscillations initially. Both the initial oscillation frequency, as well as the final asymptotic value depends on the choice of the frequency of the individual harmonic oscillators.}
    \label{fig:scalcon}
\end{figure*}

\section{Entanglement evolution in de-Sitter through scalar exchange} \label{entanglement_dS}

\begin{figure*}
    \centering
    \includegraphics[width = 15cm, height = 9.5cm]{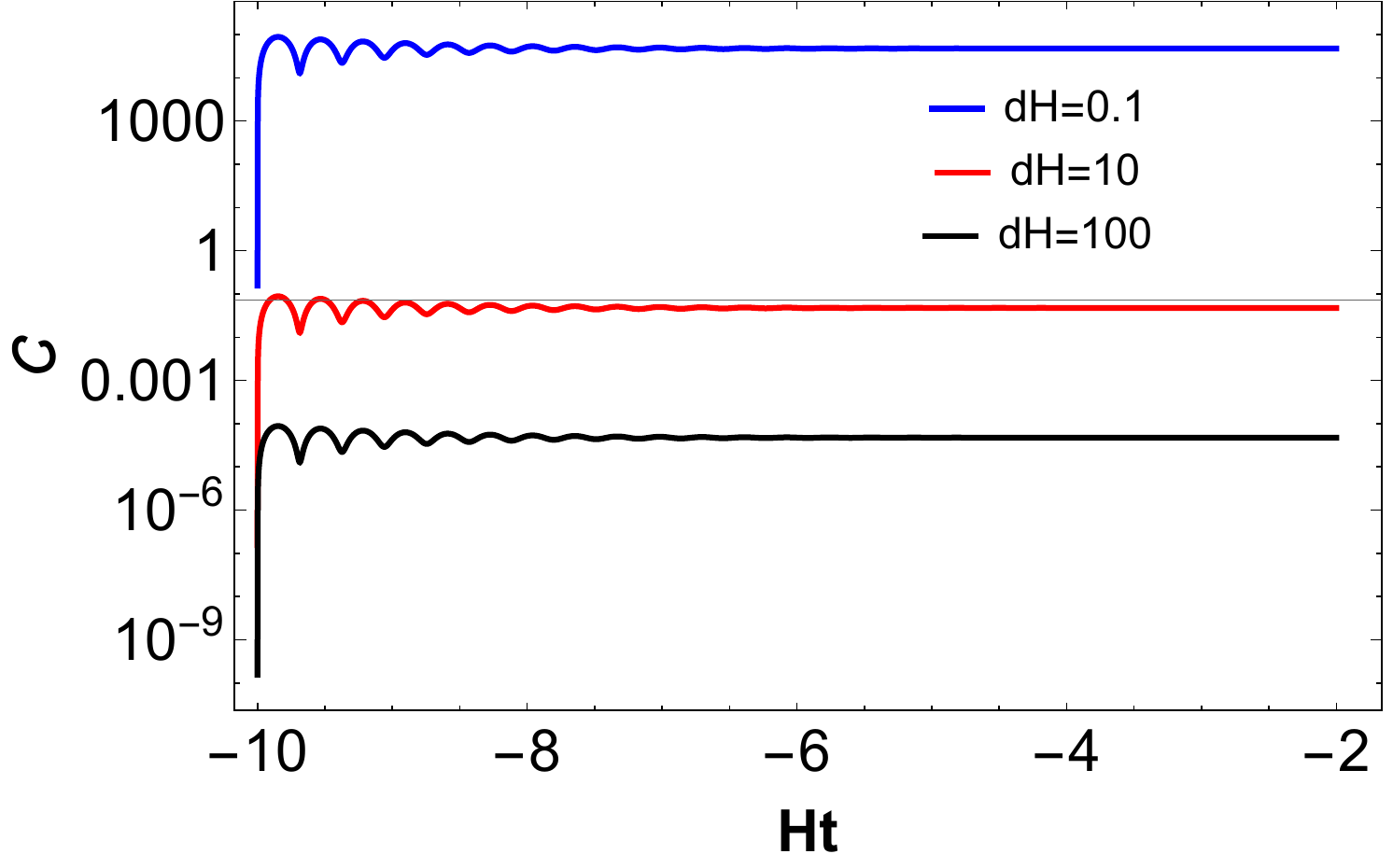}
    \caption{Evolution of the concurrence with time has been presented for three different choices of the dimensionless coordinate distance $dH$ and a fixed frequency for the harmonic oscillators, namely $(\Omega/H)=10$. As evident, the oscillation frequency of the concurrence, at the initial time, remains same in all the plots, suggesting that the initial frequency is proportional to $(\Omega/H)$. While the asymptotic value of the concurrence decreasing with increasing the distance between the oscillators.}
    \label{dvar}
\end{figure*}

To calculate the mixing coefficient between the two harmonic oscillators in dS spacetime, starting from an initial vacuum state, interacting through the Hamiltonian $\hat{H}_{\rm AB}^{(t)}$, we require the matrix elements of the interaction Hamiltonian. For that reason, we binomially expand the denominator, which following \ref{setup1}, requires expressing $|\hat{\bf x}_{\rm A}-\hat{\bf x}_{\rm B}|=d+(\delta \hat{x}_{\rm A}-\delta \hat{x}_{\rm B})$. With this substitution and subsequent binomial expansion, we get the following expression for the interaction Hamiltonian 
\begin{align}\label{interactionHamilton}
&\hat{H}_{\rm AB~(int)}=-\frac{\sqrt{3}m\mathcal{G}^{2}\hbar}{8\pi\Omega a(t)d^3} \exp \left(-|x_{0i}|a(t)dH\right)
\nonumber
\\
&\times\Big[1+a(t)|x_{0i}|dH+a(t)^{2}|x_{0i}|^{2}d^{2}H^{2}\Big]\left(\hat{a}^{\dagger}+\hat{a}\right)\left(\hat{b}^{\dagger}+\hat{b}\right)\,.
\end{align}
Here, $\Omega$ is the oscillation frequency of the harmonic oscillators. In order to find the entanglement between these two harmonic oscillators, we start with the vacuum initial state for the two harmonic oscillators, such that $|\Psi\rangle_{\rm i}=|0\rangle_{\rm A} |0\rangle_{\rm B}$, and since the Hamiltonian has one $\hat{a}^{\dagger}$ and one $\hat{b}^{\dagger}$, it follows that the final state must have the following form,
\begin{align}
|\Psi \rangle_{\rm f} = \frac{1}{\sqrt{1+|C_{11}(t)|^2}}\left(|0\rangle_{\rm A} |0\rangle_{\rm B} + C_{11}(t) |1\rangle_{\rm A}|1\rangle_{\rm B}\right)\,,
\end{align}
where, the time dependent excitation factor $C_{11}(t)$ can be obtained from \ref{gen_excite} and \ref{interactionHamilton} as, 
\begin{align}
&C_{11}(t)=-\frac{i}{\hbar}\int_{-\infty}^t dt'\,\langle 1|_{\rm B}\langle 1|_{\rm A}\hat{H}_{\rm AB~(int)}|0\rangle_{\rm A}|0\rangle_{\rm B}\,e^{2i \omega t'} 
\nonumber 
\\
&=\frac{imG_{\phi}}{\Omega d^{3}} \int_{-\infty}^{t}dt'\, e^{-Ht'+2i\Omega t'} \exp \left(-|x_{0i}|dHe^{Ht'}\right)
\nonumber
\\
&\quad \times \Big[1+e^{Ht'}|x_{0i}|dH+e^{2Ht'}|x_{0i}|^{2}d^{2}H^{2}\Big]\,,
\label{29}
\end{align}
with $G_{\phi}=(\sqrt{3}\mathcal{G}^{2}/8\pi)$. The density matrix associated with the `A' harmonic oscillator is given by \ref{densityA}, and given the form of the final state of the harmonic oscillators, takes the following form, $\hat{\rho}_{\rm A}=\textrm{diag.}(1,|C_{11}|^{2})$, except for the normalization factor. Therefore, from \ref{concurrence}, we can obtain the concurrence as,
\begin{equation}
C(t)=\frac{2|C_{11}(t)|}{1+|C_{11}(t)|^{2}} \approx 2|C_{11}(t)|\,, 
\label{30}
\end{equation}
where, we have used the result that the value of the quantity $|C_{11}(t)|$ will be proportional to $\mathcal{G}^{2}$, which is assumed to be small. Thus combining all of these results, the re-scaled concurrence between the two harmonic oscillator reads,
\begin{align}
&\mathcal{C}(t)\equiv\frac{C(t)}{mG_{\phi}H^{2}}=\frac{2}{(\Omega/H)(dH)^{3}}
\nonumber
\\
&\times \Bigg|\int_{-\infty}^{t}dt'\,e^{-Ht'+2i\Omega t'} \exp \left(-|x_{0i}|dHe^{Ht'}\right)
\nonumber
\\
&\quad \times \Big[1+e^{Ht'}|x_{0i}|dH+e^{2Ht'}|x_{0i}|^{2}d^{2}H^{2}\Big]\Bigg|\,.
\end{align}
As evident from the previous discussion, it follows that in the dS spacetime, the concurrence is time dependent, implying that the two harmonic oscillators will not only be entangled, but their entanglement will also change with time. The evolution of which has been plotted in \ref{fig:scalcon} and \ref{dvar} for different choices of the dimensionless combination $(\Omega/H)$, with a fixed value of $dH$ and vice versa, respectively. As evident from both the plots, the entanglement for all of them shows initial oscillations, while, ultimately these oscillations saturates to some asymptotic values. The oscillation frequency of the concurrence in the earlier times is proportional to the oscillation frequency $(\Omega/H)$ of the oscillators, as evident from both \ref{fig:scalcon} and \ref{dvar}, and as the frequency of the oscillator increases, the initial oscillation becomes more rapid. Further, from both the plots, it is clear that the entanglement saturates at late times in a dS background and the asymptotic value of the concurrence depends on both the frequency $(\Omega/H)$ and the distance $dH$. Note that, for our universe, the dS phase lasted for a finite amount of time, and hence the lower limit of the integration is not exactly to be set at $-\infty$, rather at some finite value. Finally, let us consider the $H\to 0$ limit, i.e., in flat spacetime, the concurrence turns out to be $\approx (2mG_{\phi}/\Omega d^{3})\sin(\Omega t/2)$. Note that this result is also consistent with \cite{Chakraborty:2023kel}.


\subsection{Asymptotic value of the entanglement}

\begin{figure}
    \centering
    \includegraphics[scale=0.4]{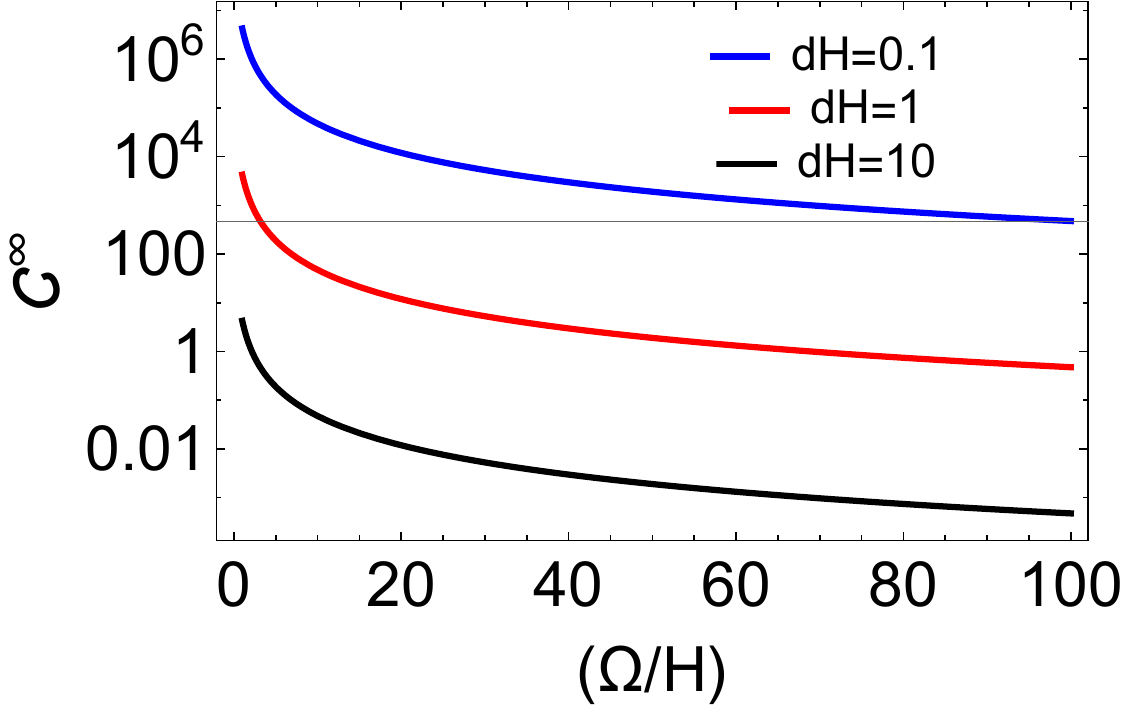}
    \caption{The asymptotic value of concurrence has been plotted against the frequency of the harmonic oscillators, for three different choices of the distance between harmonic oscillators.}
    \label{asympFreq}
\end{figure}

The asymptotic value of the entanglement between the two harmonic oscillators in a dS background is given by the quantity $C_{11}$ in the $t\to \infty$ limit, which boils down to the following integral:
\begin{align}
C_{11}^{\infty}=\frac{imG_{\phi}}{\Omega d^{3}} \int_{-\infty}^{\infty}dt'\, e^{-Ht'+2i\Omega t'} \exp \left(-|x_{0i}|dHe^{Ht'}\right)\,,
\label{asympc11}
\end{align}
where, we have ignored the corrections present in \ref{29}, as they are highly subdominant compared to the leading order exponential term inside the integral. In particular, by performing a change of variable, starting from $u=|x_{0i}|dHe^{Ht'}$, the above asymptotic limit becomes, 
\begin{align}\label{31}
C_{11}^{\infty}=\frac{imG_{\phi}}{\Omega d^{3}}\frac{(|x_{0i}|dH)^{2-\frac{2i\Omega}{H}}}{|x_{0i}|dH^{2}}\Gamma\left(-1+\frac{2i\Omega}{H}\right)\,,
\end{align}
and hence the asymptotic expression for the concurrence becomes 
\begin{align}
\mathcal{C}^{\infty}=\frac{|x_{0i}|}{\Omega (Hd)^{2}} \left[\frac{(2\pi H/\Omega)}{\sinh(2\pi \Omega /H)}\left(1+\frac{4\Omega^{2}}{H^{2}}\right)\right]^{1/2}~.
\end{align}
Thus, the asymptotic value of concurrence is non-zero and dependent on the frequency of the oscillators, as well as the distance between the oscillators. For small values of $(\Omega/H)$, the concurrence grows as, $(\Omega/H)^{-2}$, while for large values of the frequency of the Harmonic oscillators, we have $\mathcal{C}^{\infty}\sim \sqrt{H/\Omega}\exp(-\pi\Omega/H)$. Thus for larger values of $\Omega$ we have an exponential decay of the asymptotic value for the concurrence. This is consistent with \ref{fig:scalcon} and also with \ref{asympFreq}. Similarly, the asymptotic value of the concurrence decreases with increasing distance between the oscillators as $C^{\infty}\sim (dH)^{-2}$. This is evident from \ref{dvar}. Here also, since the dS phase lasted for a finite duration of time only, it follows that the asymptotic value will involve an incomplete Gamma function, whose behaviour we have plotted in \ref{asympFreq}. In summary, for harmonic oscillators, oscillating at the same order as the expansion of the universe ($\Omega\sim H$) and with coordinate distance comparable to the Hubble horizon ($d\sim H^{-1}$), the entanglement between the oscillators is large. While, either for harmonic oscillators oscillating much faster than the expansion of the universe ($\Omega\gg H$), or, for oscillators separated by a coordinate distance larger than the Hubble horizon ($d\gg H^{-1}$), the entanglement between the oscillators is non-zero, but small. 

The generalisation of the above result by including graviton exchange between the oscillators can be done straightforwardly. First of all, the graviton exchange between these oscillators, at the non-relativistic level, in flat spacetime, will introduce the same $|\hat{\bf{x}}_{\rm A}-\hat{\bf{x}}_{\rm B}|^{-1}$ term, as that of the massless scalar field in the flat spacetime. In the dS spacetime, the only difference will be the existence of an exponential piece $\exp(-|x_{0i}|a(t)dH)$ in the interaction Hamiltonian, leading to a rescaled concurrence, where $G_{\phi}\to G_{\phi}+G_{\rm N}$. All of the behaviour of the concurrence, as presented above, e.g., the initial oscillatory behaviour, as well as the asymptotic behaviour, remains the same, with simple rescaling by Newton's gravitational constant. Thus, graviton and scalar exchange between these oscillators will lead to identical evolution of concurrence/entanglement in a dS universe, see~\cite{MUKHANOV1992203,Ashoorioon:2012kh}.

\section{Conclusion}

Entanglement between two massive harmonic oscillators, resulting from the exchange of a graviton and a massless scalar, in scalar-tensor theories of gravity, can provide insights into the quantum nature of gravity and the possibility of the existence of a fifth force. Here, we have explored such opportunities in the context of an expanding universe and have demonstrated that non-zero as well as significant entanglement can exist between such oscillators, interacting in a de Sitter background, via massless scalar exchange. In particular, if the oscillators are within the Hubble horizon and oscillate at a rate comparable to the universe's expansion rate, the entanglement is large and non-zero at late times. For oscillators outside the Hubble horizon and oscillating at a rate much faster than the universe's expansion rate, the entanglement is non-zero but negligible. Even though we have shown this for the exchange of a massless scalar, as argued above, the same holds for graviton exchange as well.

Our results may have a small bearing on the inflationary epoch of cosmological expansion, where the background geometry is described by the de-Sitter metric in a scalar-tensor theory of gravity, and the harmonic oscillators can be thought of as individual modes of the fluctuations of the inflaton field. The above analysis demonstrates that without other sources of decoherence, the low-frequency modes within the Hubble horizon may retain a significant entanglement between them. However, the observed modes from the Cosmic Microwave Background correspond to high-frequency modes, for which the gravitational as well as scalar interaction between the modes leads to a final entanglement that is very small and hence effectively undetectable.

Thus, if we can measure a non-trivial entanglement or correlation between two points in the sky, essentially originating from the inflationary epoch, it may provide an indirect avenue to test the quantum nature of gravity, the fifth force, as well as the quantum-to-classical transition of the universe during cosmological expansion.
\section*{Acknowledgements}

Research of S.C. is supported by the MATRICS Grant (MTR/2023/000049) and the Core Research Grant (CRG/2023/000934) from the SERB, ANRF, Government of India. S.C. also thanks IUCAA and ICTS for their associateship program, which has also helped at various stages of this work. A.M.'s research is funded in part by the Gordon and Betty Moore Foundation through Grant GBMF12328, DOI 10.37807/GBMF12328. This material is based upon work supported by Alfred P. Sloan Foundation under Grant No. G-2023-21130.

\bibliographystyle{apsrev4-1}
\bibliography{Ref}
\end{document}